\documentclass{elsart5pag}

\usepackage{natbib}

\usepackage{graphicx}

\usepackage{amssymb}

\begin{document}

\begin{frontmatter}



\title{Complexity Threshold for Functioning Directed Networks in Damage Size Distribution}


\author{Andrzej Gecow\thanksref{homead}}
\ead{gecow@op.pl}
\thanks[homead]{01-923 Warsaw, Boguslawskiego 4/76, Poland, gecow@op.pl}
\address{Centre for Ecological Research Polish Academy of Sciences,\\
M. Konopnickiej 1, Dziekanow Lesny, 05-092 Lomianki, Poland}

\begin{abstract}

A certain complexity threshold is proposed which defines the  term `complex network' for RSN, e.g. Kauffman networks with $s\geq 2$ - more than two equally probable state variants. Such Kauffman networks are no longer Boolean networks. RSN are different than RWN and RNS. This article is the second  one of three steps in description of `structural tendencies' 
which are an effect of adaptive evolution of complex RSN.  This complexity threshold is based on the appearances of chaotic features of a network during its random growth and disappearance of small network effects. Distribution of damage size (after small disturbance) measured in a fraction of damaged nodes, or in number of damaged external outputs and degree of chaos is investigated using simulation. It is done during growth (up to $N$=4000 nodes) for different: network types (including scale-free), numbers of node inputs ($K=2$,3,4, fixed for a network) and numbers of signal variants ($s=2$,3,4,16). In this distribution two peaks emerge and in-between them there appears an area of zero frequency - this is the best practical criterion of complexity threshold found in the investigation. No critical points are found in the area of emerging complexity. A special simplified algorithm (`reversed-annealed') is used which omits the problem of circular attractors. The investigated `transition' to chaos in respect to $N$ is different from the known phase transition near $K=2$ for $s=2$.

\end{abstract}

\end{frontmatter}

\section{Introduction}
\label{ch.1}

\subsection{Complexity threshold - what we are looking for}
\label{ch.1.1}

There is no commonly used definition of complexity but terms `complex' network, and `complex system' are used often and they are important because they describe one of the most dynamically developing discipline of science. Complexity has many different aspects, each one leading to its own definition of complexity. Each of those definitions can be adequate in a particular range of phenomena. Probably one common definition does not exist. There are as many different definitions \citep{Peliti88, Jost04, Yamano} as different aspects and applications of this notion. On the Internet a `Complexity\_Zoo' can be found: ``There are now 489 classes and counting!''. Typically together with a definition, a certain measure of complexity is also proposed,  e.g., Solomonov-Kolmogorov-Chaitin algorithmic complexity \citep{Ming97,Gell-Mann95} or simply number of elements of system \citep{ooKauf} is used. In Ref. \citep{Olbrich} two measures of statistical complexity are tested when the system size is increased: are the results compatible with intuitive requirements for complexity measures in three special cases: (i) adding one element, (ii) adding an independent system and (iii) adding a perfect copy of the original system.

However,  the terms `complex network' and `complex system'  are understood as two state alternative: complex or not complex. Such understanding needs some threshold to be defined, over which a system or network is complex but below it - not complex. 
Our intuition connects this threshold to our ability to predict the behaviour of a given system. If it needs too many well measured parameters and too many of interaction and conditions which should be taken under consideration, then we name it as too `complex' to be solved.  But such `definition' is too subjective.
In order not to be a subjective criterion, this threshold must be an effect of a particular objective phenomenon, e.g. in self-organising criticality \citep{Bak96,Bak88} after some parameters reach a critical value, the system will spontaneously exhibit behaviour characterized by power laws or in scale-free networks \citep{ss,Dorogovtsev03}, above a critical network size, the average length of the path between two nodes will not change with the growth of the network.

The term `big system' was replaced by `complex network' in widespread use. Our threshold should be therefore a threshold in network size upon which effects of small networks (`finite size effects') disappear in practice and `large network' is a good approximation.
Continuing with the subjective suggestion we can convert it to a particular formulation in the network language: If similar initial states of system typically lead to similar effects, then in average for one state which should be predicted, with high probability we can predict the effect simply by comparing initial states to the known ones. Such a case is named `structural stability' in Ref. \citep{ooKauf}, I prefer not to name it `complex' despite the fact that network can be large. However, if similar initial states lead to very different effects, then the prediction is not so easy and it seems complex. Sensitivity to initial conditions is the most important property of `chaos'. It can be observed by watching the effects of a small initial difference of two functioning systems. Typically \citep{Jan94} two identical systems are taken, in one of them a small `disturbance' is introduced, and differences called `damage' in states of network nodes during consecutive steps of function of both networks (i.e.: `damage spreading') are watched. Networks in which damage typically grows are named `chaotic' \citep{ooKauf}. This notion of chaos is a little different from the typically used and well defined \citep{Schuster84} notion for continuous functions. It considers finite networks with finite (however - large) number of states. Damage cannot grow infinitely but it reaches an equilibrium level instead. However, in this range of applications it is undoubtedly the same phenomenon. 

Our subjective meaning of a `complex network' appears to be similar or equal to the term `chaotic network' and is connected to the number of network elements $N$. Let us check if there exists in degree of chaos some threshold-like phenomenon when number $N$ of network elements grows. In other words: is there a `transition' to chaos (from order) or something similar during network growth? We can treat this transition as the complexity threshold. It should objectively indicate a particular value of $N$, but this value may depend on other parameters of network. The notion: `degree of chaos' is also unclear and can be understood in different ways. Clarifying this picture is the main goal of this paper. 
In the effect we found that investigated phenomenon is not a critical one - chaotic properties emerge smoothly, however in relatively short stage of growth, and the main practical criterion depends a little on number of events. It is an effect of critical point (percolation) which was crossed for much smaller network.

Such an approach bases on the assumption that living objects (which we are going to describe) are chaotic much more than the Kauffman's conception `life on the edge of chaos' expects, because homeostasis based on negative feedbacks are there underestimated \citep{agec}.

As was mentioned in the beginning, each complexity definition or criterion is designed for a particular application. This approach can (and does) give us a certain definition of complexity, but it is (like others) not a general one. It does not consider e.g. ordered networks, which can also be complex in some respects for some applications. Such defined complexity and its threshold are connected to the mechanism of structural tendencies which appear as statistical effects of adaptive evolution of complex (in this meaning) functioning networks but they do not appear for small networks. This criterion is needed for definition of range of their appearances. Structural tendencies are a main goal of my approach which start from Ref \citep{agec} where RSN and `reversed annealed' algorithm are introduced.
Complexity threshold is now discussed, and on this base I will next investigate structural tendencies. I hope that the proposed notion and criterion of complexity of networks and systems fulfil well most of the expectations when term `complex network' is used.  

\subsection{Contents of the paper in brief }
\label{ch.1.2}

This article is a next step after Ref. \citep{agec} where RSN - Random, ($s\geq 2$) equally probable Signal variants Network was introduced, `reversed annealed' algorithm was described and living systems were shifted on chaotic shore of Kauffman's liquid area near the edge of chaos. This two articles create a base for investigation of structural tendencies shortly described in Ref. \citep{dgec}.
Therefore in the second section (ch.2) brief reminding of Ref. \citep{agec} can be found,  necessary for understanding of the current paper. However, without reading Ref. \citep{agec} a lot of less important details could remain unclear. One of the most important theme (negative feedbacks and position of living systems on order-chaos axis) discussed in Ref. \citep{agec} is shifted here to ch.4.5 `Important interpretative remarks'.

Next, new (relatively to the first article) additional assumptions are defined in ch.3. 
1- In the first article autonomous networks are considered, now connection to environment is introduced and networks become not autonomous. External inputs are fixed, but on the external outputs of network certain damage is observed. In the next step fitness will be defined on external network outputs.
2- In adaptive evolution investigation in the next step of my approach not only node additions but also removing of nodes must be considered. Removing of nodes creates new phenomena which strongly influence network features. They significantly change $P(k)$ `node degree distribution' because node with $k=0$ (without outputs) are created. Therefore two new network types are added to investigated set of network types. These are networks $sf$ (scale-free similar to SFRBN) and $ss$ (single scale similar to EFRBN) with 30\% of random removals of nodes which, for the purpose of this paper,  are named $sh$ and $si$ respectively.

Chapter ch.4 describes simulations and their results. Mainly damage size distribution is watched during network growth in two variants - fraction of damaged nodes and number of damaged outputs for different network types and parameters: $s$ - number of equally probable signal variants and $K$ - number of node input fixed for a particular network.
Discussion of precision of obtained results and causes of their dispersion is interesting. It uncovers large natural fluctuation of investigated processes and leads to experiment which is a base of name `reversed-annealed' for used algorithm. This experiment could not have been described in Ref. \citep{agec} together with algorithm. 
Degree of chaos in dependence especially on network type and differences of both types of damage size are discussed in this chapter. 
In all four considered dependencies a critical point doesn't occur in region of `complexity threshold', however, matured chaotic features emerge and effects of small network  practically disappear.

Section ch.5 is dedicated for interpretation, explanation and intuition. It helps in understanding the obtained results of simulation. Especially the depth in network is defined there, which differ places inside network. It is the smallest path from node to network external outputs.

\section{Reminding of Effects of First Step}
\label{ch.2}

This article is a next step after \citep{agec} creating base for structural tendencies. As a starting point we consider  ``quenched'' - deterministic dynamics of the Boolean networks \citep{ooKauf}. 

The first step has introduced a few important elements necessary for understanding this paper, especially RSN (Random Signal Network), which takes there a significant part of the article - their description should not be fully repeated in any next paper - only the most important elements will be recapitulated in short here.
   
\subsection{Main set of parameters ($s$, $K$, $N$, $k$, $d$)}
\label{ch.2.1}

There was introduced a parameter $s$ - number of equally probable variants of signal which can be greater than two and this defines RSN. Its role in degree of chaos (in wide meaning) was discussed for different network types as well as a comparison to the influence of $K$ parameter (number of inputs per node, fixed for particular network in this approach). 

As opposed to most works using Kauffman networks I do not agree with the ``life at the edge of chaos'' hypothesis in Kauffman's form. I believe that the Darwinian mechanism is much more powerful than this hypothesis assumes and is the source of main part of order and stability found in living objects. 
I estimate that chaotic networks are typically more adequate for description of real living objects and the case $s>2$ cannot be neglected, but it causes that Kauffman networks become something more than Boolean ones and RBN is not  an adequate name any more.
In the current article the degree of chaos is the main investigated parameter, now its relation to the number of system elements $N$ will be examined. However in Ref. \citep{agec} the `degree of chaos' is not a defined parameter yet. There were discussed a few phenomena connected to chaos which can be used to assess the level of chaos in a network. One of these phenomena will be formally named `degree of chaos' $c$ in ch.2.3 (in similar way like `degree of order' in Ref. \citep{agec})
but it is only one of aspects of general notion of  `degree of chaos'.

An important effect of values of $s$ other than the typical one ($s=2$) is higher equilibrium level of damage $d$ (- fraction of nodes with damaged state). In ch.2 in Ref. \citep{agec} a theoretical expectation (``annealed approximation'' model) independent of network $type$, (in Kauffman network range) described for $s=2$ in Ref. \citep{Derrida86}, (also in Ref. \citep{ooKauf} p.200 fig.5.8, or in Ref. \citep{Iguchi07} fig.B.2) were expanded for higher $s$. For ordered networks, the levels of damage (not predicted using such method) are much lower which can be one of the characteristics used to distinguish between not chaotic and chaotic networks. Functions of nodes are assumed to be fully random, however, they are virtually not used in Ref. \citep{agec} and here in simplified simulations.

\subsection{Set of network ($types= er, sf, ss, aa, ak$) and coefficient $w$}
\label{ch.2.2}

The research will be continued on the same basic set of network types defined in Ref. \citep{agec} in details. In short: Kauffman networks are the most adequate to describe adaptive systems because they function which produces an effect which can be assessed and improved. 
The function formula of Kauffman network gives us a useful ability to differentiate $k$ (- number of outgoing links of nodes, degree of node) within the network.

Kauffman and many other researchers who followed him used Erd\H{o}s-R\'enyi random networks \citep{er,Kauf69,ooKauf}, but with $s=2$ and fixed number of incoming links $K$. These networks are named RBN (Random Boolean Network), but later this name has got two meanings - it describes also a wider class of networks and the name `CRBN'\citep{Serra04A} (Classical) for Erd\H{o}s-R\'enyi random networks was proposed.
I denote it as $er$ but I use $s\geq 2$ (in range of RSN). 
I have introduced $s>2$ in Ref. \citep{agec}, (earlier in Refs. \citep{paris,hof}). Kauffman network and Boolean network were synonyms. I think that RSN (without aggregate of automata network family), RWN\citep{Luque04,Luque05} and RNS \citep{Luque97, Sole00} using more than two variants of node state should remain `Kauffman networks'.

Different types of networks differ in the distribution of node degree $P(k)$. 
Nowadays famous Barab\'asi-Albert scale-free network seems to be more adequate in most cases to describe reality \citep{sf99, sf03} than the old $er$ network. 
Using it in the Kauffman network frame I denote it as $sf$, however in Ref. \citep{Iguchi07}  where $s=2$ and $K$ is not fixed, the name SFRBN is used. Such a network was earlier investigated in Ref. \citep{Aldana03, Kauf04,Serra04A}.  
This network $type$ has a characteristic `preferential attachment' pattern for growth which leads to power law distribution of node degree.
Node degree distribution $P(k)$ for  $er$  is a bell-like curve which ends quickly for relatively small $k$.
Between these two types lies the next candidate for investigation - `single-scale'\citep{ss} network $type$ $ss$  which grows without preferences. 
In Ref. \citep{Iguchi07} it is named EFRBN (Exponential-Fluctuation Random Boolean Network) 
and is similarly considered together with SFRBN (but there $s=2$ and $K$ is not fixed).
Its node degree distribution $P(k)$ decreases faster (exponentially) and is less extreme than for $sf$ networks.

The coefficient $w=\langle k\rangle(s-1)/s$   of damage multiplication on one node is introduced in Ref. \citep{agec}, earlier in Refs. \citep{paris,hof} and similar in Ref. \citep{3Aldana}. 
It is a useful notion for understanding of damage spreading which will be used also in this paper for explanation.
To understand the coefficient $w$ in the Kauffman network where 
all $k$ outputs of a node transmit the same signal, 
we must average the changes of signals and the $k$ for lots of nodes. 
It is much simpler and more intuitive and the averaging of the change is possible in one node if $k$ is fixed and each output of a node has its own signal to transmit. 
In such a case $\langle k\rangle =k=K$ and the function value (node state, outputs) is a $K$-dimensional vector.
I have introduced such a network in Refs. \citep{hof, krab, bic, dgec, agec} where I have named it `aggregate of automata', therefore I name it here $aa$. 

For complex $aa$ network the `structural tendencies' was first 
investigated and they occur very strong \citep{krab,bic}. 
For comparison to the Kauffman networks described above, 
the $ak$ network $type$ similar to $aa$ was defined
which follows the Kauffman network rule (one signal on all outputs), 
but $k=K$ remains fixed.
However, networks with flexible $k$ are more rich in different phenomena and mechanisms.

\subsection{Special algorithm, range of parameters $s$ and $K$ in simulation and two peaks}
\label{ch.2.3}

In Ref. \citep{agec} above set ( $sf,ss,er,ak,aa$) of networks was simulated for different parameters $s$ ($s=2$,3,4) and $K$ ($K=2$,3,4) near transition from chaos to order. Note, that $s$ is the number of equally probable signal variants defining RSN. From full set of combinations the case $s=2$, $K=2$ was removed because it is an extreme one - it always leads to ordered network, but these values are the lowest sensible of these parameters. 
In Refs. \citep{Aldana03,Kauf04,Iguchi07} the range $1< \langle k\rangle \leq 2$ is investigated as the most interesting one, however, they mainly consider ordered phase and the point of phase transition to chaos. In such a network there must appear nodes with one input and one output but using fixed $K=1$ network becomes extreme and I neglect such case.

For all other combinations (higher values of parameters $s$ and $K$) chaos is expected. Higher value of each parameter leads to higher level of chaos, e.g. degree of chaos $c$ measured using fraction of events in right peak in distribution of damage size. This peak is directly connected to the large damage avalanche which is the main feature of chaos in opposition to left peak which contains ordered behaviour events of fade-out. These two peaks and a pass between them will be the main theme of investigation of this paper. 

However, there is also the $N$ parameter (number of nodes) whose changes were not discussed in Ref. \citep{agec}. The values used were high $N$=2000 and $N$=3000 and for used parameters ($type,s,K$) the results are practically identical. For small values of $N$ damage has no space to convert itself into an avalanche and chaotic features cannot express. Between such small and high values of $N$ a threshold of chaos must appear and this is the theme of this paper. As it will be shown, for $s=2$ and $K=4$ the threshold appears for about $N$=1200 for $ss$ and $N$=300 for $er$, but $N$=4000 is too few for $sf$. In Ref. \citep{Iguchi07} maximal $N$ for $K=4$ is 150, far from threshold even for $er$. Our threshold is different than the one considered in Ref. \citep{Derrida86, ooKauf} $K=2$ for CRBN ($er$) or found in Ref. \citep{Iguchi07} in range $1.4< \langle k\rangle < 1.7$ for SFRBN.

The specific case $s=2$ $K\leq 2$ (excluded above) cannot be simulated using special simplified algorithm prepared for investigations of chaotic networks developed in Ref. \citep{agec} and here.
The same simplified algorithm dedicated for simulation of statistical damage spreading and described in details in Ref. \citep{agec} will be used in current paper. Two important notions resulting from the algorithm definition should be reminded - `real-' and `pseudo-fadeout' of damage. Pseudo-fadeout corresponds to reaching of an equilibrium level of damage. Process ends because damaged node is calculated only one time. 
In the classic approach starting from \citep{Kauf69} the length and number of circular attractors are investigated. They are interpreted e.g. as cell types. The described simplified algorithm is designed to omit the problem of circular attractors and allows to define fitness on network outputs (see next chapter). In the full quenched model such a particular output does not exist and fitness should be defined on the attractors of outputs, 
however, the problem of fitness appears only for target structural tendencies. Now, in a middle step, only damage size $L$ on outputs will be defined using this algorithm feature. The length and number of circular attractors are connected to the equilibrium level of damage in some unknown way and are not investigated directly in described simulations.

The basic conclusion of simulation made in Ref. \citep{agec} which is important for our discussion was the indication of distribution $P(d )$ where $d$ is a damage size (- number of damaged nodes divided by $N$ - number of all nodes) when damage fades out (independently of the way of fadeout). In this distribution there are two peaks; one near zero containing real fadeout and second on the right (mentioned above) containing pseudo-fadeout which describes the equilibrium level of damage.

The existence of this right peak is a feature of chaotic network. The algorithm gives correct statistical values in such a case, however we are interested in the earlier period - in appearance of chaos and this second peak. This area is depicted using this algorithm with some expected inaccuracy, but the threshold of chaos should be also a threshold of fully correct effects. For our mostly qualitative discussion in this area such inaccuracy in not important but should be remembered. 
For simulation in Ref. \citep{agec} and in current paper particular functions are not used, such simplification is useful for statistical summarising and averaging of results. However, in later investigation of structural tendencies functions are necessary. Simplification removes certain elements of: `deterministic' and `quenched' features of simulated model, however, it does not reach the `annealed' level. Later there will be used a case where structure of connection of node is stable but all states of nodes are randomly changed - looks like `reversed-annealed'  case and from this our algorithm takes the name.
 
Other conclusion from simulations described in Ref. \citep{agec} useful for current discussion is the sequence of network types in degree of chaos $c$ for the same parameters $s$ and $K$ which is shown in fig.6 in Ref. \citep{agec} (however, in parameter $r$ - degree of order). These investigations are expanded in ch.4.2 (see fig.5).The less chaotic $type$ is $sf$, which is also concluded by 
\citet{Iguchi07}. It is because small a part of nodes (named hubs) collects the major part of links which leads to much smaller local coefficient $w$ for the remaining part of nodes which defines this area as ordered. Consequently, damage often really fades out before it reach the first hub which can help it to explode and convert into avalanche. More chaotic $type$ is $ss$, then $er$, but only the $er$ network contains nodes with $k=0$, which have strong influence on network behaviour. On the `chaotic' end of Kauffman networks there is the $ak$ network. Only $aa$ network is more chaotic than $ak$ in this method of measurement of chaos.

\section{New Additional Assumptions}
\label{ch.3}
 
\subsection{Connection to environment: $L$ damaged of $m$ outputs}
\label{ch.3.1}

Following ref. \citep{ooKauf} the size of damage  $d\in \langle 0,1\rangle $ is measured as the fraction of nodes with damaged output state in the all nodes of system. 
This parameter, however, is typically hard to observe for real systems. 
Adaptation process concerns interactions between system and its environment. 
If we are going to describe such a process, then damage should be observed outside the system, on its external outputs. 
However, network with outputs is no longer an autonomous network like the ones considered from Ref. \citep{Kauf69} up till Ref. \citep{Iguchi07}. Some links are special - they are connected to environment which is another, special `node' which does not transmit damage (in the first approximation) unlike all the remaining ones. Damage fades out on the outputs like on a node with $k=0$, therefore the dynamics of damage $d$ should be a little bit different depending on the proportion of output size and network size. Environment as an objectively special node can be used for the indication of nodes' place in a network, which without such special node generally have no objective point of reference. The main task of this special node in adaptation process is a fitness calculation and Darwinian elimination of some network changes.

The simplest definition of damage size on system outputs is: the number $L$ of damaged output signals. 
For large networks with feedbacks it is applicable only using a simplified algorithm described in Ref. \citep{agec}. It omits the problem of circular attractors.
Formally, $L$ is a Hamming distance of system output signal vectors between a control system and a damaged one. Practically, using our algorithm, it is distance between system output before and after damage simulation. 
For simulations the system has a fixed number $m=64$ of output signals which means that $L\in \langle 0,m \rangle $. 
We can expect, that distributions $P(d)$ and $P(L)$ should be similar. 
In fact, asymptotic values (for `matured systems'): $dmx$ of $d$ and $Lmx$ of $L$ are simply connected: $dmx=Lmx/m$  
but such a connection is not true for smaller systems and $L$ is smaller than expected. 
Note, that number of output signals $m$ is constant and much smaller than the growing number $N$ of nodes in the network, 
which must influence on the statistical parameters and their precision.  

\subsection{ Removing of node, $k=0$, networks $sh$ and $si$ }
\label{ch.3.2}

\begin{figure}[b]
\begin{center}
\includegraphics[width=8cm]{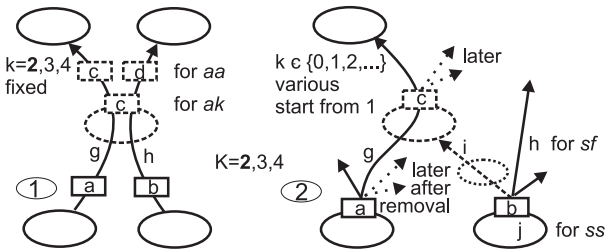}
\end{center}
\caption{Changeability patterns for $aa$ and $ak$ (1), $sh,sf,si$ and $ss$
network (2) depicted for $K=2$. In  case of addition of a new node to the 
network,  links $g$ and $h$ are drawn. 
Node $j$ is drawn instead of link $h$ for $ss$ and $si$. 
For $K>2$ additional inputs are constructed like the right ones.
The $ak$ network is maintained as $aa$ but there is only one, 
common output signal $c$.
An extension in comparison to fig.3 in Ref. \citep{agec} 
is removal of node which only needs a draw of a node to remove.
Main moves are the same as for addition but in opposite sequence, 
however events which occur after the addition change situation 
for $ss$ and $sf$.
Removal can create case $k=0$: node added on link $i$ 
remains a $k=0$ node while removing.
The outgoing links, which were added after addition of this node to the network, 
are moved to the start node of link $g$. 
This lack of symmetry causes changes in distributions $P(k)$ 
and other features of a network, 
therefore networks $sf$ and $ss$ with removals of node are different 
than without and are named $sh$ and $si$ respectively.}

\label{fig:1}     
\end{figure}

In the ch.3.3 in Ref. \citep{agec} the importance of fraction of nodes with $k<2$ for level of chaos was discussed. In the $sf$ and $ss$ networks which are built using only additions of nodes, as in Ref. \citep{agec}, only $k=1$ can be found in range $k<2$ (i.e. no node has $k=0$). In the $er$ network $k=0$ also appears and this leads to different behaviour. However, in the target model of adaptive evolution and structural tendencies using only addition of node during network grow will be insufficient, there also removing of nodes is needed. 
Random removal of a node needs to draw a node only from the nodes constituting the network (see fig.1). 
Each node should have equal probability to be chosen.

Pattern of  disconnection of node should be the same but in opposite direction to connection when adding, however, if removing happens not directly after addition, situation can change and such simple assumption will be insufficient. Such a case appears for $sf$ and $ss$ networks when $k$ of removed node can be different than - just after addition and, which is especially interesting, when on the right input link a new node was added. During the removal, this new node loses its output link and may become a $k=0$ node. Nodes with $k=0$ and other nodes connected to them, which have no way to external outputs are called `blind' nodes. The existence of `blind' nodes in the network is one of the biggest and the most interesting problems especially for the modelling of adaptation. The importance and complexity of this problem is similar to the problem of feedbacks. 
The set of investigated networks in the step described in this paper leading to structural tendencies should contain networks built with removing of nodes, which means $sf$ and $ss$ should contain $k=0$. Such networks are different than the typical $sf$ and $ss$, also because removals change node degree distribution, therefore networks built with 30\% of removals of nodes and 70\% of additions get other names - $sh$ for modified $sf$ and $si$ for $ss$ with removals. (I keep the second letter in two letter names unique,  which is useful in practice, especially when there is no place for two letters in the figure.)

\section{Simulation}
\label{ch.4}
 
\subsection{Description of the experiment}
\label{ch.4.1} 

Fig.2 shows effects of simulations in distributions $P(d|N)$ and $P(L|N)$ for different network types and their parameters $s$ and $K$. 
As in Ref. \citep{agec} these parameters are denoted in form: two (or one) letters of network $type$, followed by $s$ and $K$ parameters in number separated by a comma. 
All combinations of 7 network types, $s=2$,3,4,16 and $K=2$,3,4 except $s,K$ = 2,2 and 16,4 were simulated. This is $7*(4*3-2)=70$ cases.

\begin{figure*}
\begin{center}
\includegraphics[width=18.2cm]{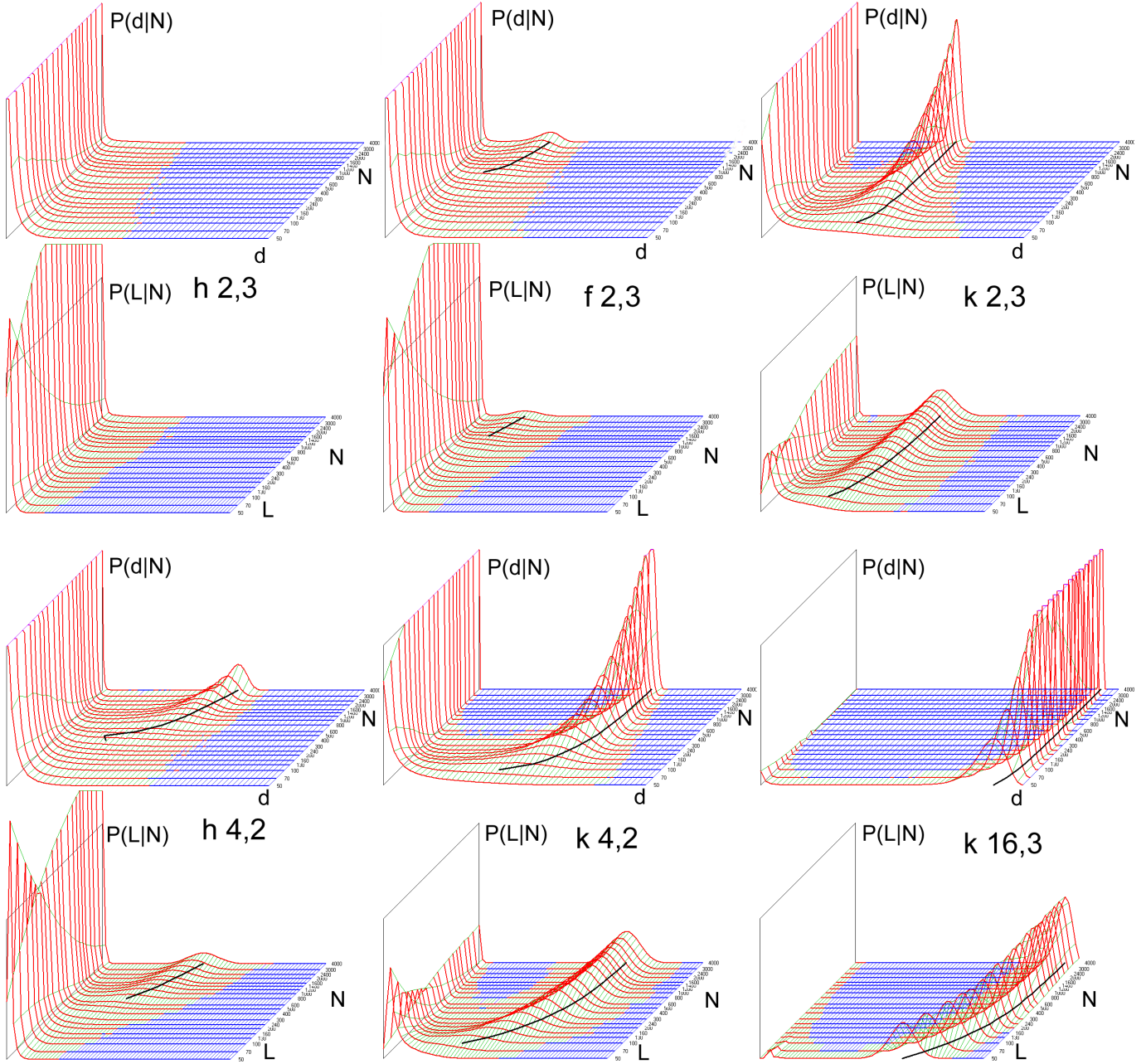}
\end{center}
\caption{The main view of evolution of distributions of damage sizes $d$ (upper) and $L$ (lower) during network growth for different network $type$, $s$, $K$. 
Selected examples of the basic results of simulations showing full spectrum of main features from the less chaotic $sh$ 2,3 to the most chaotic $ak$ 16,3.  $N$ axis is in logarithmic scale. Blue indicates zero frequency. Vertical projection (black line) of maximum of right peak on the `sea level' plane ($P$=0) is the equilibrium level of damage. These lines from all simulation cases are collected $d(N)$ in fig.3 and $L(N)$. in fig.4. The proportion of the volume of right peak to all the events of damage initiation is the chaos degree $c$ depicted in fig.5.}

\label{fig:2}     
\end{figure*}

Networks grow during the simulation (except $er$ which is drawn each time) and at the levels of $N$ = 50, 70, 100, 130, 160, 200, 240, 300, 400, 500, 600, 800, 1000, 1200, 1400, 1600, 2000, 2400, 3000, 4000 damage is simulated. As in Ref. \citep{agec} all remaining variants of state of each node are used in turn as damage initiation. When they all are used, the network is grown to the next level of $N$. Results of a few networks with the same parameters are summarised.  In effect of such a method a number of events for a small $N$ will be much smaller than for last (far greater) values. For this reason the number of events is assumed one for each levels of $N$ and when for bigger $N$ it is achieved then next network ends their growth on lover levels of $N$. This number of events is for $s$= 2, 3, 4, 16 equal respectively 400, 320, 240 and 300 thousands, whereas for $sh$, $sf$, $si$, $ss$ 2,3 it is 800 000. It is different than for more chaotic networks to obtain a similar exactness of the right peak. 
When $s$=16, then there are 15 remaining states and to obtain 300000 events for a network containing $N$=4000 nodes it is enough to build 5 networks. 
As it will be shown later, it is a relatively small number and a large but interesting fluctuation was obtained (fig.6).  
Distribution $P(d|N)$ ( more exact - number of events of fade out on $d*N$ nodes) has different ($N$) points of `$d$' for different $N$. For all cases there are used fixed 200 points to make comparison available, but such operation must cause some discontinuities in the beginning of the first, left peak. 

\subsection{Depiction of the results}
\label{ch.4.2}

Each case of 70 cases described above is depicted in 2 types of damage size ($L$ and $d$). In fig.2 only 6 cases are shown, but they depict well all the basic phenomena. This is the main view of evolution process of distribution of damage size $d$ and $L$ during network growth. In details of this process the threshold of chaos will be searched as parameter of network maturation and complexity.

\begin{figure}[b]
\begin{center}
\includegraphics[width=8.8cm]{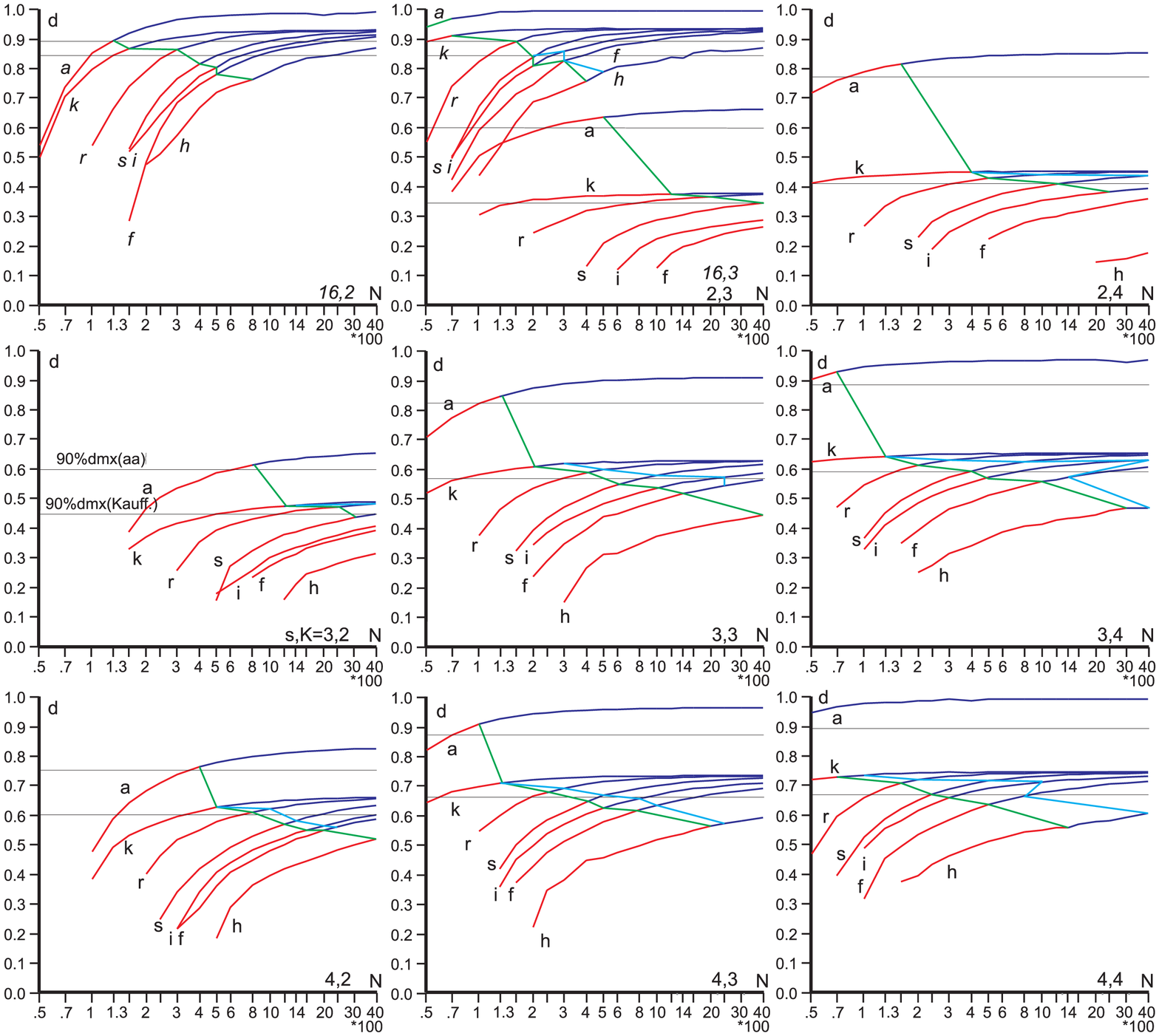}
\end{center}
\caption{Collection of all simulated cases $d(N)$ - equilibrium level of damage size in the network.
Type of network is denoted by the second letter only, for $s$=16 in italics. Red part of the curve - before zero appearance between peaks. 90\% of asymptotes $dmx$ are depicted in back horizontal line.}

\label{fig:3}     
\end{figure}

In the plane parallel to the paper a distribution $P(d)$ or $P(L)$ is shown for particular network size $N$, which grows depicted in logarithmic scale in depth behind the paper. More exactly, these are frequencies obtained as simulation results. 
For non-zero values of $P$ (as frequency) red colour is used, for zero value - blue. For particular values of $L$ or $d$ the $P$ values for different $N$ are connected by lines, which are blue if both connected values of frequency ($P$) are zero and green otherwise.
This useful method easily depicts dynamics of chaos appearance as nice landscape with the important area of blue `bay' where exact zero between peaks occurs. Right peak emerges from the tail of the left peak in left part of damage size (small damage) and in the first period of network growth it moves to the right, but later it turns smoothly and its position becomes stable, drawing an asymptotical line (black one) to parallel line to $N$ axis. It is vertical projection of maximum of right peak on `sea level' ($P=0$). These lines from all cases of simulations are collected in fig.3 for $P(d|N)$ and in fig.4 for $P(L|N)$. 
Between peaks the pass forms whose minimum goes down and reach zero frequency (forms the blue bay). In fig.3-5, 7 the part of each curve with this zero frequency on the left of right peak is indicated by colour (blue or green).

\begin{figure}[b]
\begin{center}
\includegraphics[width=8.8cm]{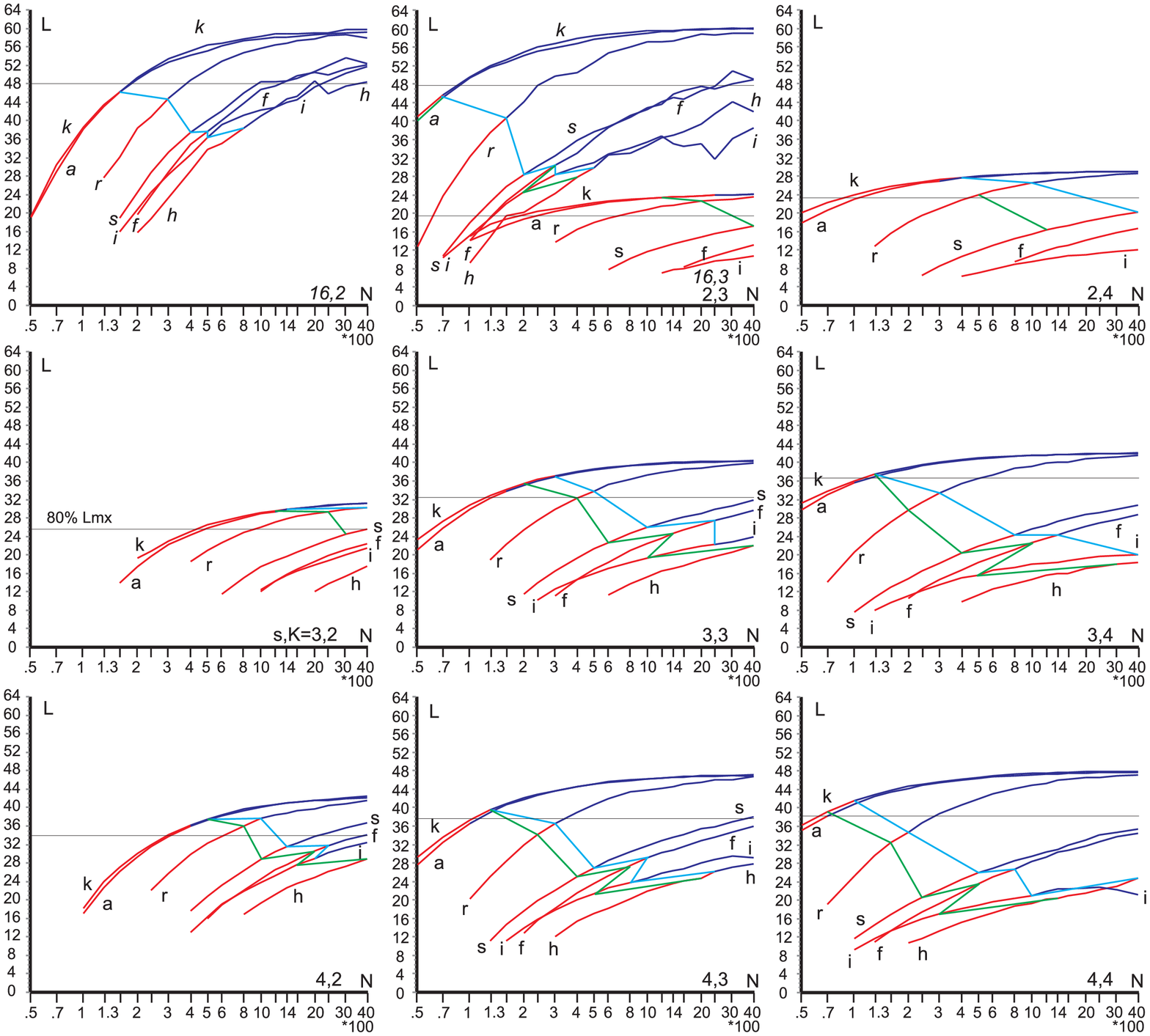}
\end{center}
\caption{Collection of all simulated cases $L(N)$ - equilibrium level of damage size on the external outputs. 
Type of network is denoted by the second letter only, for $s$=16 in italics. Red part of the curve - before zero appearance between peaks. 80\% of asymptotes $Lmx$ are depicted in back horizontal line.}

\label{fig:4}     
\end{figure}

The position of the line parallel to $N$ axis to which projection of maximum asymptotically goes (dmx for $d$ and Lmx for $L$) can be calculated theoretically using $s$ and $K$, independently (except $aa$ for $d$) of the network $type$, as was described and shown in `Derrida plot' in fig.1 in Ref. \citep{agec} which is an extension of the method \citep{Derrida86,ooKauf} using annealed approximation for large networks. It is the level of damage equilibrium for a big chaotic network. 
If network is not big enough, then this level is not reached and the main feature of chaos - damage avalanche - cannot fully develop.
It can be used as a criterion of chaos threshold, similarly like the occurrence of a zero-frequency area between peaks which definitively separates peaks. 

Note, left peak contains real fadeouts, which are `ordered behaviour' of the network, whereas the right peak contains damage which reach their equilibrium level. 
When this level is high, near theoretical for big chaotic network (dmx or Lmx), then such a network should be treated as fully chaotic.

Preliminary researches \citep{dgec} made mainly on $s=4$ and $K=2$ suggest that 90\% of dmx, 80\% of Lmx  and zero area occurrence, which all occurred in similar network size $N$, may be sensible criteria for the threshold of chaos. Unfortunately for other parameters this coincidence does not appear which was indicated in Ref. \citep{dgec} for all other $sf$ cases investigated in those simulations: 4,3  16,2 and 64,2.
Current, much wider investigations can show how similar the results given by these three criteria of matured chaos threshold are.
   
Right peak of $P(L)$ is much wider than the one for $P(d)$ and this width is approximately constant, but for $P(d)$ the right peak becomes narrower and higher when $N$ grows. This is a typical statistical effect: the number $m$ - maximum of $L$, is constant and much smaller than the growing number $N$ of nodes which may be damaged. For higher $N$ the right peak of $P(d|N)$ is so high that it must be cut which is visible in fig.2 for $ak$ 4,2 and $ak$ 16,3. 
Left peak also often must be cut.

In fig.2 we start from the case $sh$ 2,3 which in the investigated area of $N$ is always ordered without any trace of chaos. This is the less chaotic network as we found later, and $s,K$=2,3 is the less chaotic combination. Next, a little bit more chaotic is the $sf$ network which for 2,3 exhibits first steps of emergence of right peak. The $ak$ network is the most chaotic of Kauffman networks  - for 2,3 it exhibits quite an advanced stage of chaos emergence.
It contains a large area of zero between peaks for $P(d)$ and a small yet but not negligible area of zero for $P(L)$. The equilibrium level of damage for $s=2$ is low which makes the emergence of zero difficult. 
In the lower row in fig.2 the case of 4,2 for extreme network types $sh$ and $ak$ is shown. Network $sh$ reaches zero between peaks for $P(d)$. This $type$ of network reaches a state similar to $ak$ 4,2 for 16,3 - it is very similar to $si$ 16,3 shown in fig.6.1. The lower right part of fig.2 depicts the most chaotic simulated combination of Kauffman networks parameters - $ak$ 16,3.

In this paper the emergence of chaos is investigated. A really chaotic network can be considered `mature' both in the sense of network complexity and in the sense that its small network effects (finite size effects) are practically negligible. Degree of chaos (in general sense) is treated as a good parameter describing maturation level of network - means: level of possibility to practically neglect effects of small network, and complexity level. 
We are looking for the threshold of chaotic features and its criteria. 
Degree of nearness to theoretical level of damage equilibrium for a big network is a feature of chaos connected to the ability of damage avalanche to reach its full power. 
Appearance of zero between peaks clearly separates ordered and chaotic events of damage after damage initiation. It makes our algorithm exact and allows to clearly interpret its results. However, it does not inform directly about the level of chaos in the right peak, which means damage equilibrium can be far from the theoretical value for big chaotic network. Correlations with high level of this equilibrium will be discussed. This appearance of zero in frequencies is not an effect of low number of events - there probability is really very small. 

Network can exhibit approximately full chaotic feature, the damage avalanche reaches the theoretical equilibrium level, but it can happen in a small part of events of damage initiation, in the second larger part the damage can really fade out. It depicts degree of chaos $c$ - the fraction of all events which constitute the right, chaotic peak.
Similarly to the appearance of zero area, this method shows another aspect of chaos than we are looking for and its correlations with high level of damage equilibrium should be discussed. 
In the critical point of percolation a `large cluster' appears, but near this point it is typically still not really large and it only appears occasionally. For higher N this cluster appears more frequently, which is indicated by parameter $c$. 
This measure of degree of chaos $c$ for full set of simulated cases is shown in fig.5. 
For higher N the `large cluster' also becomes a larger fraction of all nodes of the network, but here a statistical limitation appears which depends on network type. It is the maximal average cluster appearing for very large network which defines an asymptote. 
More precisely - higher N cause that the `average large cluster' becomes a larger fraction of maximal `average large cluster'. 
Note, that the size of cluster is measured as fraction of N, not in number of nodes in the cluster.
This phenomenon is observed in position of the second peak and its distance to asymptote. It was described above in other terms, see fig. 3. 
Note, that all these measures lack one important aspect (see ch.4.5) for living objects which is an effect of adaptive evolution.

\begin{figure}[b]
\begin{center}
\includegraphics[width=8.8cm]{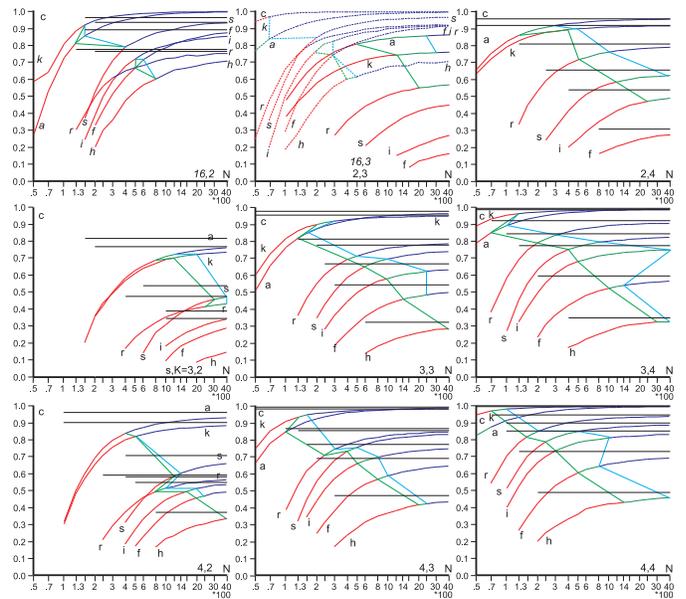}
\end{center}
\caption{Collection of all simulated cases $c$ - degree of chaos: the fraction of all damage initiations which cause damage avalanches. Asymptotes are depicted in back horizontal line.
Type of network is denoted by the second letter only, for $s$=16 in italics. Red part of the curve - before zero appearance between peaks, green - after zero appearance for $d(N)$ and before for $L(N)$. }

\label{fig:5}     
\end{figure}

\subsection{Precision and causes of dispersion, reversed-annealed simulation}
\label{ch.4.3}
We are interested in the right peak, its separation level from the left peak and position of its maximum. Gaussian-like shape of this peak suggests connection with statistical source and a question appears which features of this peak are an effect of number of simulation events and not of investigated features of this peak. 
We have denoted above that the width of this peak depends on parameters $m$ for $L$ and $N$ for $d$, but is this the only source of this wideness?

\begin{figure}[b]
\begin{center}
\includegraphics[width=8.8cm]{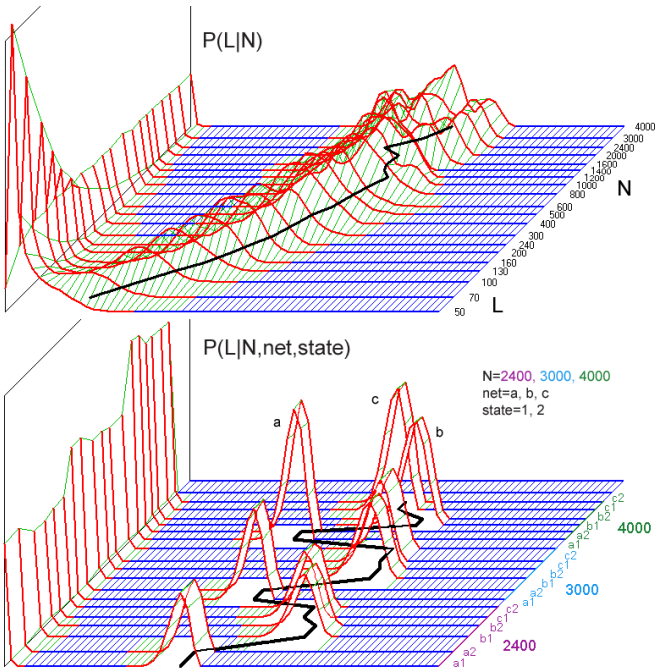}
\end{center}
\caption{Individual structure influence on the right peak width. (1) - Fluctuation in $P(L)$ distribution observed at the end of growth in $si$ 16,3 case. For each $N$ there are 300000 events but there are only 5 networks for $N$=4000 and $s$=16.
(2) - Extreme example of result of additional simulation for analysis of fluctuation causes: 3 networks ($a,b,c$) on $N$=2400, 3000, 4000 measured two times, second time after a random change of all node states. Value of shift is larger (for $s$=16) than the width of peak which is exactly repeatable (two consecutive cases). Stable character of deviation for particular network during its three steps of growth is visible.}

\label{fig:6}     
\end{figure}

Earlier we noted that for $s$=16 where only 5 networks were used for the last $N$=4000, large fluctuations are observed (fig.6.1). This observation we will discuss first, because its cause is clear. Each particular network has its own particular individual structure which should cause a deviation from the average behaviour. 

To check this hypothesis of cause of observed peak deformation three networks were simulated for $N$=2400, 3000 and 4000. To confirm that the structure is the cause, for each $N$ level two damage measurements were performed - one with node states as they were obtained during network growth and the second one with new random states of all nodes. Fig.6.2 shows one of the more extreme examples of effects of this experiment. As was expected, effects of changing node states and original states are identical - this  method resembles annealed model \citep{Derrida86,ooKauf} but reversed: in the original annealed model, the states are kept and the structure is changed. Note, for simulations in this and previous \citep{agec} papers we do not need to use particular functions for nodes. 

A question appears: are these structural features, which cause the observed deviations, stable during next levels $N$ of network growth? Fig.6.2 suggests such stability. Indeed, such a stability appears typically and the distances of peak maxims are  larger than the standard deviation of peaks for one network.

As can be observed in fig.6.2, the peaks' width is small (because $s$=16 is large) and the distance between peaks is larger which is a basic cause of visible deformation in fig.6.1 for last $N$.
To estimate the influence of this individual deviation of structure on the total peak width an average standard deviation for one network peak was measured and compared to the standard deviation of total peak for many networks as depicted in fig.2. For network $ak$ in both: $d$ and $L$ there are no differences between width of total and individual peaks. For $er$ and $ss$ also there are no visible differences, but only for $d$. However, for $L$ standard deviations of total peak are from 20\% to 100\% larger. For $ss$, $si$, $sf$ and $sh$ network types the wideness of total and individual peaks and their absolute differences consecutively grows, but it depend on $K$ and $s$ parameters. These differences relatively and absolutely are the biggest for $sh$ 16,3.

If we assume that each node or output signal has equal probability to be damaged and we use as this probability theoretical $d$ (ch.2 in Ref. \citep{agec}) or the $d$ obtained in simulation, then the right peak is described by a binomial distribution with known parameters. 
We can then compare its standard deviation to the one obtained in simulation experiment. Such simple model works too well - for $L$ experimental data are in range of error (10\%) but for $K=3$ and $K=4$ obtained standard deviation is a little bit lower than theoretical. 
It also approximately well explains values of standard deviation for $d$ and $ak$ network $type$, however, for $K<4$ experimental values are greater than the theoretical ones up to two times for $s,K$=4,2. This discrepancy grows for more chaotic networks and reaches 4 times higher value for $sh$ 4,2. For these comparisons only $s=4$ and $s$=16 were used because for lower $s$ the errors in data are too large.

In conclusion - the wideness of right peak has two known causes. The first one is theoretical and results from $m$ or $N$ parameter and the equilibrium level of damage which depend on $s$ and $K$. This first cause does not depend on the number of experimental events and is a feature of the network case (combination of $type$, $s$, $K$). This aspect can be used for theoretical approximation of slope of probability in area of `zero occurrence'. The second cause is the individual structure of particular networks, it can be minimized using more networks for determining the maximum point. The above analysis suggests that for more chaotic networks (which in this article were given names starting from letter `$s$') there is some significant unknown mechanism of dispersion. The above simple theory does not explain the observed dependency on $K$ for damage type $d$.  For damage type $L$ (which should be an effect of $d$) it probably covers more complex dependencies. 

\subsection{Analysis of simulation results}
\label{ch.4.4}

Full information for analysis is contained in fig.3-5, 7. There are collected equilibrium levels of damage size measured in $d$ (fig.3) and in $L$ (fig.4) for different network sizes $N$ and parameters $s$ and $K$. This dependency is a line obtained as a projection of the right peak maximum onto the plane $d*N$ or $L*N$. Red colour depicts its first period, typically not yet matured, without zero between peaks. When zero occurs the colour changes to blue in fig.3 and fig.4. Points of zero appearance for different network types are connected using green or blue line (for more complex areas they are omitted). Levels of 90\% of dmx ($d$ equilibrium for large network) and 80\% of Lmx for $L$ are shown. For $d$ they are different for Kauffman networks and for $type=aa$. 

In fig.5 the degree of chaos $c$ as fraction of events in right peak is depicted. Here red is used if in both $d$ and $L$ measures there is no zero area to the left of the right peak. 
This zero occurs firstly for $P(d)$ and from this point the line is green, later zero occurs for $P(L)$ and from this point the line is blue.

Parameter $s$ changes from 2 (upper chart) to 4 (lower chart), however there are no cases 2,2 and 16,4 and $s$=16 is placed together (to save place) with $s=2$  in the first line which is a little bit not easy to read, sorry. To make this part more readable, italic font is used for $s$=16. Parameter $K$ changes from 2 to 4 going from left to right. For each case of $s,K$ all seven network types are depicted (if they all have right peak). Networks are described using only one (second) letter of the name. 

In fig.3 where $d(N)$ is collected all network types always create the same sequence: the lowest is $sh$, then $sf$, $si$, $ss$, $er$, $ak$ and the topmost one is $aa$ network with different asymptotic level. The lines never cross each other. 
The picture of $L(N)$ is similar (fig.4), but now $ak$ and $aa$ lie close to each other ($aa$ systematically a little bit lower). The $type$ $si$ typically crosses $sf$ in the beginning and their sequence on their right end is opposite than for $d(N)$ and for $c(N)$ in fig.5 where $er$ $type$ line crosses the line of $ss$ for $s=2$ and the lines $ak$ and $aa$ cross or change their relative position. 
In their main features, these three pictures are similar.

Appearances of zero typically occur for $d$ and $L$ close to each other, separated by one step of $N$ therefore it can be treated as one criterion of chaos. However it also happens, e.g. for $si$ 4,4,  that the occurrences are further away from each other. For $L$ one zero case was enough to indicate zero appearance but for $d$ 5 cases of zero was used (fewer than 5 zero cases were neglected). 
The steps of $N$ are too large to determine this place exactly but obtained data are enough for analysis. 
For set of networks $type$ with the same combination $s,K$, the  lines of zero appearance are typically sloped. It means that there are no coincidences between the criteria 80\% of Lmx or 90\% of dmx and these criteria of chaos threshold give different results. 
Looking through the full set of data in form of fig.2 we can conclude that appearance of zero looks correctly in any case: the right peak is never too small and always has a `matured' symmetrical shape.

Criteria based on the certain fraction of asymptotical value of damage equilibrium have a convincing definition and indicated point with error which can be determined, however they do not work in accordance with the level of visual `maturation' stage of network. 
For smaller $s$ and more chaotic network, e.g. $ak$ 2,3 or 2,4 or even 3,2 (fig.3 crossing of 90\%$dmx$ or fig.4 - 80\%$Lmx$)  the right peak is small (fig.2 $k 2,3$) and between peaks there is only a shallow pass whereas less chaotic networks are typically (fig.3 and 4) even for $N$=4000 out of this criterion as not chaotic, however, they often (even $sh$ 4,2 in fig.2) have a `matured' shape and zero between peaks in this point. 

This unexpected conclusion is explained in fig.5 where the degree of chaos $c$ measured as part of events in the right peak is depicted. `Less chaotic networks' by fraction of asymptotical value indeed typically have significantly lower $c$. It suggests that a large part of each of these networks is ordered, but the chaotic part of network is `matured' and avalanche of damage in this part realises most of its growth potential. It means, that for such an avalanche the effective network size is smaller. 

If we depict the distance to asymptote ($dmx$ or $Lmx$ also in approximation for $c(N)$) in log-log diagram, then we obtain straight lines. They depict `finite size effects'. It indicates that these relationships are scale-free and we cannot expect in them any special values of N. Particular point of assumed exactness is a point of practical approximation, not a physical phenomenon, however, real phenomenon of  sought `complexity threshold' exist and a method to describe it should be found. Such assumed exactness does not seem  satisfactory.

For degree of chaos $c(N)$ distance to asymptote on log-log chart is nearly straight but getting a bit convex. If we neglect the first few points about which we know that they are not sufficiently precise due to algorithm features, it gets straighter. Assuming that the relationship is linear, we obtain an asymptote which can only lie too high. These asymptotes are individual for each particular parameter vector $type,s,K$, they are shown in fig.5 in black horizontal lines. They mean that even a very large network is not lacking an ordered behaviour which occurs in even high faction of damage initiations. Classic Derrida annealed approximation neglects this important feature. Second conclusion is that $type$ of network is a crucial parameter defining degree of chaos of large networks.

Such a view is in accordance with differences of $sf$ and $sh$ network or $ss$ and $si$ - nodes with $k=0$ moves them to lower degree of chaos in similar way as nodes with $k=1$.
The degree of chaos describes the probability that in effect of a damage initiation an avalanche occurs but the equilibrium level of damage size $d$ and $L$ considers only cases where avalanche actually occurs. 

For our task of indicating the maturation point we can remark that zero occurrence is denoted on $c(N)$ curves typically on the end of their fast growth where the curve takes a significant turn towards the horizontal direction. On the log-log diagrams of distance $c$ to its asymptote lines connecting points of zero occurrence are less slopped than such lines for $d$ and $L$. This observation increases the importance of `zero occurrence criterion'.

Let's come back to the problem of proper region for adaptive evolution. In Ref. \citep{agec} living systems were shifted into chaotic area, however, they still remain within the boundary of Kauffman's liquid region. There\citep{agec} was considered `degree of order' $r$ shown in fig.6.2 which equals $r = 1-c$.
Now we have estimated asymptotes for different $type,s$ and $K$ and we can try to estimate the boundary of `liquid region' in chaotic direction i.e. for higher $s,K$. There is not enough data for a more precise  approximation, for $s=3$ and 4 there are only 3 points of $K$, but for higher $K$ changes of asymptote levels are significantly smaller and we can expect that even for high $K$ the `degree of order' $r$ for $type = sh,sf,si,ss$ is not to be neglected. It can even mean that e.g. for the important $sf$ network there is no shore in chaotic direction.
I think that a practical shore exist, but adaptive evolution has its region of real $r>0$ even for highly chaotic networks (chaotic in other sense than $c$).

\subsection{Important interpretative remarks}
\label{ch.4.5}

Here I must stress important interpretational reason: In this paper adaptive process is not assumed, however, applications of this investigations for living objects are planned. 
Splitting a set of effects of small disturbance into only two subsets - real (random) fadeout and damage avalanche ending in equilibrium is a large simplification for living objects. Adaptive evolution collecting `biological information', whatever it means in details, chooses special networks which are called `homeostatic'. This important term is used in literature in different meaning: usually it means negative feedbacks which keep system in a particular range with respect to many parameters. Such features were named `ultrastability' by 
\citet{ooKauf} probably follow Ashby. Lots of such mechanisms are typically met in living objects. Life is often saw as a process of collecting negative feedbacks e.g.\citep{Benio01,Benio05}. It is probably an overestimation of this aspect, however, the condensation of negative feedbacks in living objects is extremely high on all levels of their structure and function. This concentration is an effect of the pressure of the adaptive condition. In our simulation there is no such pressure and no its effects. Our model is far  too simple to investigate an appearance of homeostatic mechanisms. However, for such a `abstractive' more adequate model if there were such a pressure, then we should observe a very large and important fraction of events with initial small `semi-random' disturbance, when the damage later fades out because such homeostatic mechanisms keep control of it. Such disturbances are `semi-random' because the system met them earlier during the adaptive evolution and `knows' a method to prevent their effects. (More exactly - not a system understood as all its ancestors met such reasons but sisters of these ancestors which are eliminated.) GRNs - Gene Regulatory Networks based on 
\citet{ Banzhaf03} work are such more adequate models, where negative feedbacks emerge.

There are feedbacks in our networks, but they are random and they do not perform homeostatic functions which can only be an effect of the adaptive condition. From the outside of the system, the effect of homeostatic  mechanism seems to be similar to real fadeout, however, the latter one is only a random effect and our estimation of its frequency is based on such assumption (that is random, homeostatic reaction is not a random). 

\citet{ooKauf} uses the term `homeostatic' in a different meaning: his `homeostatic stability' is a spontaneous resistance of system to disturbance which is a property of `ordered' type of system.  Adaptation founds such system type which is the main effect of adaptation for Kauffman. This is the essence of his `life on the edge of chaos' concept. Damage avalanches on great scale are impossible in ordered systems, they should quickly really fade out or stay on a low level of equilibrium. It is exactly these properties that we investigate in our simulations, however, I believe that the main effect of adaptive evolution is not the Kaffman's `homeostatic stability' but the negative feedbacks. Kauffman's gene regulatory model \citep{Kauf71, Serra07} works mostly in ordered regime. This model fits the experimental data well \citep{Wagner01, Serra04, Serrajtb07} and has become a strange exception from my general estimation that living objects are chaotic, however, it does not model homeostasis based on negative feedbacks (which covers majority of observed biological stability as I estimate above).

If we go this far into interpretation, then the other similar aspects should be also raised for completeness. Damage avalanche may have only one interpretation for living objects - it is death and elimination. Therefore their adaptive evolution can only be based on those random changes,  whose initiated damage either really fades-out, in range of cases investigated here, or is neutralized using a homeostatic feature, but they are out of the scope of cases examined here.
If we neglect such an important and large part of mechanisms and cases connected to them, then are these investigation adequate for the description of living objects? - Yes and no. `Yes', because we investigate a part of a set of mechanisms which work in networks describing living objects. `No' in these problems in which we are going to conclude that we have found all main causes or mechanisms of some phenomena - for these rest of problems we should investigate the part neglected here. Such an approach assumes that the interference of these effects can be neglected in the first approximation. In Ref. \citep{agec} this theme is discussed more in detail. There living systems were interpretatively shifted from ordered to chaotic shore of Kauffman's liquid area near edge of chaos.

\subsection{ Differences of $L$ and $d$ damage sizes }
\label{ch.4.6}

In ch.3.1 we remark, that we do not see reasons why the distributions $P(d)$ and $P(L)$ should be different. Indeed it is observed that $dmx$ and $Lmx$ are simply connected: $dmx=Lmx/m$.  However, such a connection is not true for smaller systems and $L$ is smaller than expected. This departure is shown in fig.7 where $L/m/d$ is depicted. This picture has similar character to all the pictures previously described and shown in fig 3: $d$, fig.4: $L$ and fig.5: $c$ - they all show certain phenomena which asymptotically disappear for large networks.

\begin{figure}[b]
\begin{center}
\includegraphics[width=8.8cm]{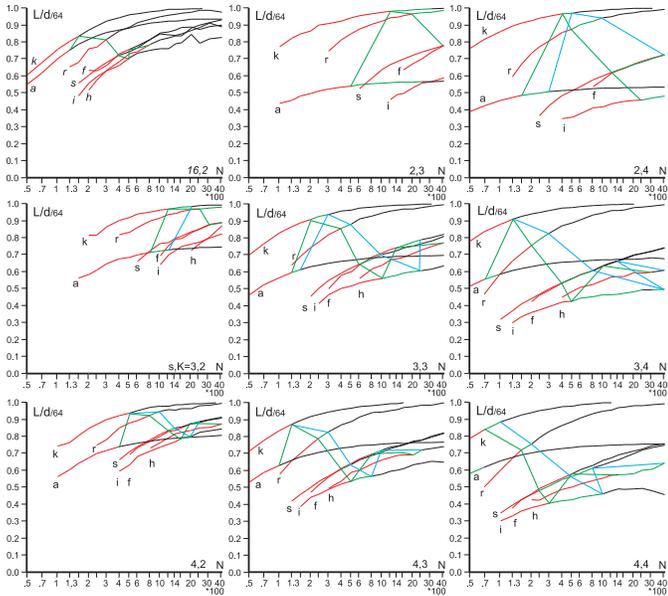}
\end{center}
\caption{Collection of all (without 16,3) simulated cases $L/m/d$ - departure of expectation $L=d*m$.
Type of network is denoted by the second letter only, for $s$=16 in italics but case 16,3 is omitted. Red part of the curve - before zero appearance between peaks, green - before zero appearance for $L(N)$. }

\label{fig:7}     
\end{figure}

This phenomenon has an unknown cause. Intuitively, the asymptotic value should be equal to one, however, it is far from this value for $N$=4000 for many cases. (For $aa$ network $type$ it is not one, but it is understandable for this case - $d$ is higher because node output states are $K$-dimensional.) It means that certain phenomena of small network should not be neglected even for $N$=4000 for networks whose names (in our terminology) start with `$s$'  ($ss$, $si$, $sf$, $sh$). The networks in this set are more important because of their applications. They have a rule of growth which leads to creation of hubs - forming the long tail of node degree distribution, and to large probability of $k<2$. The remaining considered types of network $aa$, $ak$ and $er$ are useful mainly for the detection of causes of observed phenomena because of their small ($er$) - or lack  ($ak$, $aa$) of dispersion in node degree distribution or higher damage dynamics for $aa$ $type$. 

It is interesting that in fig.7 the curves for $ss$ and $sf$ are typically identical (except the very beginning) but $si$ and $sh$ are different. Here the curve for $si$ network $type$ is extremely low and flat which is difficult to understand. Without explanation of this effect we are unable to assess the meaning of zero appearance (appearance of an area with zero) for $si$ network. For small $K$ and remaining network types the `zero appearance' well indicates the beginning of area in which small network effects can be neglected. However, for larger $K$ we observe lower curves except for $ak$ and $aa$ where the curve moves little bit up as $K$ grows, and for networks:  $ss$, $sf$, $sh$  and $s=3$ \& 4, $K=3$ \& 4, even $er$ 4,4 zero appears too early. 

Summarising the search for criteria described in chapter 4, the best candidate seems to be the criterion using `appearance of zero', however, it works correctly only for networks with feedbacks, which will be developed in the next chapter. Term `zero appearance' as well as exactness level allowing to neglect small network effects should be treated as symbols of wide set of more exactly defined criterions based on these phenomena. 
It can e.g. be an appearance of value equal certain small fraction of right peak height between peaks of normalized distribution. To this can be added requirements of range of such appearance, e.g. number of points or fraction of damage range. Always, however, we must arbitrary chose certain value which makes our threshold not fully objective. It is not a useful and nice criterion but we find that in investigated range there is no better one. In effect we never obtain  an objective point criterion but fuzzy set of such points. Particular point criterion can be defined for particular necessities and we don't developed them here.

\section{Two Peaks Explanation}
\label{ch.5}
 
\subsection{The extremely simple $lw$ network and cone of influence}
\label{ch.5.1}

Two separated peaks, one for real fade out and second for pseudo fade out are an assumption of the simplified algorithm used for simulation in Ref. \citep{agec} and in this paper. Using this algorithm we investigate these two peaks and the appearance of the right one. Isn't it a tautology? 
No, however, in the area where second peak emerges this algorithm is not clear and may not produce exactly correct effects. Such effects have no clear interpretation. Therefore it is desirable to model this area in another way to understand why we should expect two peaks and why they are not present from the beginning. 
Percolation theory is a known theory which explain these two peaks. Another way for intuition for different point of view and different aspects will be shown here.

\begin{figure}[b]
\begin{center}
\includegraphics[width=8.8cm]{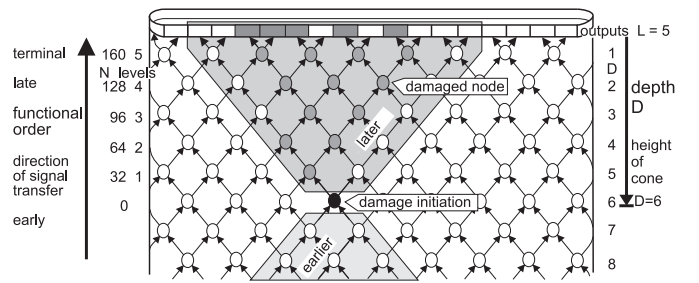}
\end{center}
\caption{Definitions of: $lw$ network, cone of influence, functional order, terms: early, late and terminal, levels and depth and their relations.}

\label{fig:8}     
\end{figure}

Let us start from the extremely simple network depicted in fig.8. Imagine that there are 32 nodes in each level. Each node has two inputs and two outputs like $aa$ network for $K=2$. Nodes from one level are connected to higher level nodes in an extremely ordered way depicted in the figure. To eliminate boundaries on the left and right they create a cylinder. Network ends at the top in external outputs of network which send signals to the environment as a result of network function. Let us name this network `$lw$' - `l' because it has clear levels and `$w$' because this letter is similar to the connection pattern. 

The $lw$ network does not contain feedbacks, it is probably the most ordered one in the sense of simplicity of connection pattern which requires an extremely small amount information for its description. Only for a similarly simple network can one draw on  paper a `cone of influence' of a given node.  The cone contains in its `later' part all nodes which can be damaged if the given one is damaged. Arrows depict the direction of signal flow - it is a directed and functioning network like all the networks of our interest (RSN) in this paper and the previous one \citep{agec}. In such a case information in form of signals flows up from the bottom - this is functional order (or sequence) which defines terms `early', `later' and `terminal' (closely near outputs, at the end of signal path). Similar phenomenon named `supremacy' is investigated in Ref. \citep{Holyst04} in more theoretical way for directed scale-free network.

Levels can be numbered starting from an indicated node, however, nodes are similar and such a method is not stable and not objective. It can be substituted by a numbering which starts from the stable outputs therefore such numbering is a depth measure - it is a useful structural substitution of the functional sequence. For a cone of influence it is the  height which affects the number of outputs which can be reached by damage initialised in the vertex of the cone. Not all of the later nodes and network outputs will be damaged - the density of filling  the cone by damage depends on the coefficient $w=k(1-s)/s$ of damage propagation described in Ref. \citep{agec} and mentioned in ch.2.2. If $w>1$ and damage does not fade out in the first few steps, then it also becomes a cone and is greater for greater cone height.  

\subsection{Damage size on outputs and damage path through network ($lw$, $lx$) without feedbacks, complexity}
\label{ch.5.2}

Let us investigate distribution of number $L$ of damaged output signals in dependency on depth $D$. As an equivalent of using the depth $D$ for the real network outputs we watch the sets of node outputs in consecutive higher levels (numbered by $H$) from initiation of damage. In fig.9.1 this process is shown for $s=4$, but levels are numbered with $N=H*32$ - number of nodes in levels from initiation of damage. Similarly to fig.2 we find here two peaks, zero in-between and the process of emergence of right peak, however, it does not yet correspond with fig.2. To obtain the same meaning as in fig.2 we must define the height of full network (which we can assume to be equal to $H$) and summarize distributions from all equally probable levels (from the beginning to the shown one $=H$ ).  The result is also shown in fig.9, but only for two heights $H$ of network (for $H,N=50,320$ for $lw$ or 50,1600 for $lx$  and 125,4000 for both). Now we also find two peaks, but in-between there cannot be zero because middle levels exist and were summarized.

\begin{figure}[b]
\begin{center}
\includegraphics[width=9cm]{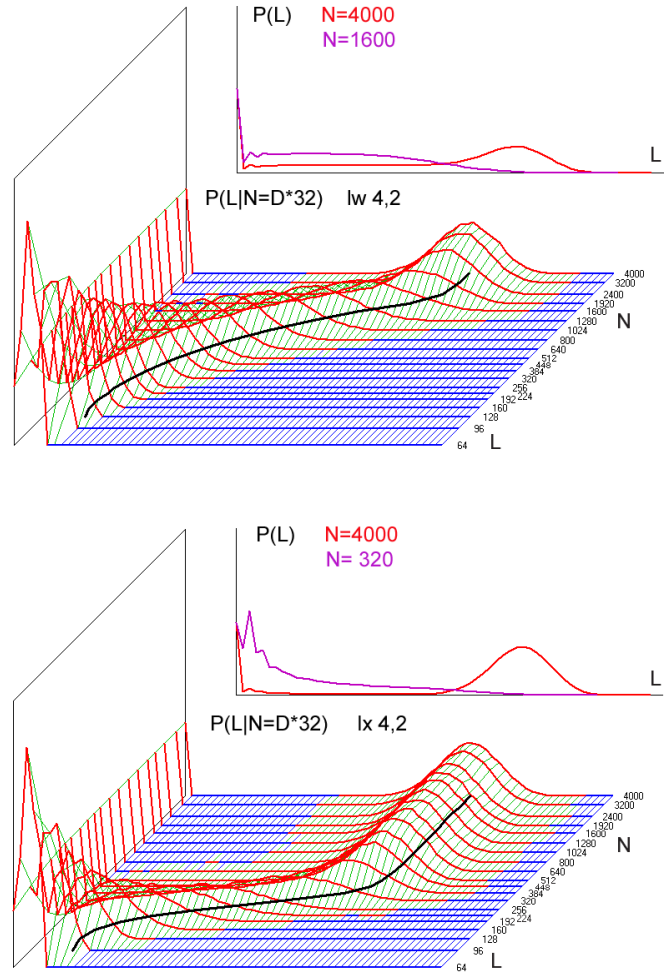}
\end{center}
\caption{Distributions of damage size $L$ for path length - levels (depth $D$) are numbered in $N=D*32$ from damage initiation for (1) $lw$ network and (2) $lx$ network. There are two peaks emerging like in fig.2, however, to obtain the same meaning as in fig.2, these distributions should be summarized. Effect of this summarizing is shown in additional diagram for $N$=1600 and $N$=4000 for $lw$ (1)  and  $N$=320 and $N$=4000 for $lx$ (2). Note that for these networks there are no feedbacks and these distributions cannot contain zero in-between peaks.}  
\label{fig:9}     
\end{figure}

Let us remove a part of `order' from the description of structure - the connection between levels are now random. We need much more information for the description of such a particular network - it is more complex. We will use the name `$lx$' for this network, fig.9.2 shows (similarly to fig.9.1) damage propagation process in consecutive damage size distributions for growing distance from the initiation point. In this case the right peak reaches its stable position much faster and the transitional stages contribute much less to the sum between peaks. 

When $s$ grows, e.g. two times from 4 to 8 then the effect is similar to an increase of complexity - we need to know more to describe particular network with states and the right peak reaches a stable position two times faster.

\subsection{Damage path through network with feedbacks}
\label{ch.5.3}

Now we should discuss the network case with feedbacks. There appears a question: how deep is a particular place of damage initiation in such a network?
This question can be translated into a question about the length of path from initiation point to outputs, however, in opposite to networks $lw$ and $lx$ such paths may be very different. 
They may be different for two reasons, first, because clear levels used in $lw$ and $lx$ are a strong assumption which must not be taken even for network without feedbacks, e.g. for $an$ network described in Ref. \citep{dgec}, second, because feedbacks make such a measure undefined.

Feedbacks are the loops in which signals can loop any times up to infinity therefore there is no the longest path, but the shortest path, which is only defined, is not well adequate. In a typical randomly constructed network there are lot of feedbacks, they are a cause of damage equilibrium state which should remain stable ad infinity. If damage passes the first period when it is small and has real probability of fade out, then it reaches the feedbacks loops and its path through network will be infinite - it correspond to an infinite depth of initiation, however, outputs may also be reached fast. Therefore in a network with feedbacks there are no middle paths and middle damage sizes placed between peaks and in this area there occurs zero frequency. This zero means that the network is so large that feedbacks are reached quickly (after a short way, frequently) enough that middle ways cannot happen.

For a network without feedbacks there remains the criterion of stable position of right peak and for practice some percent of asymptotical level should be used, however, a large flat area between peaks of small value of probability in comparison to the right peak may be taken as good approximation of zero occurrence.  

\section{Conclusion}
\label{ch.6}
 
We are looking for complexity threshold during growth of Kauffman and similar networks defined as RSN \citep{agec} (Random, equally probable Signal variants Network). An important parameter $s$ (number of equally probable signal variants) was introduced in the definition of RSN as an exclusive alternative to bias $p$ (probability of one of signal variants).

We concern ourselves with an expectation (based on interpretation) that complexity is connected to the necessity of lots of information for prediction. This leads us to the investigation of level of chaos in different aspects and to observations of a maturation process in which properties of small networks (finite size effects) disappear. Our complex network is therefore a network so large that for assumed parameters: $type$, $s$ and $K$ it exhibits chaotic features where properties of small network can be neglected. However, when and which effects can be neglected depends on lots of reasons, and among them - on our intention as well.
In order to target the application (structural tendencies) of these partial investigations we are going to observe the external outputs of network, however, we do not limit ourselves to the output. 

The main starting observation is the distribution of damage size $d$ (damaged part of nodes constituting network) or $L$ (number of damaged outputs of $m$=64 all outputs) after a small damage initiation for different network sizes $N$ (from 50 to 4000 nodes) and for different other network parameters: $type$ (Kauffman networks: $sh$ - scale-free with 30\% node removing, $sf$ - scale-free without node removing, $si$ - single-scale with 30\% node removing, $ss$- single-scale without node removing, $er$ - Erd\H{o}s-R\'enyi random network, $ak$ - fixed $k=K$; not Kauffman network: $aa$ - aggregate of automata i.e. fixed $k=K$ and $k$-dimensional output state - each output of one node has its own signal), $s$ (number of signal variants: 2,3,4,16) and $K$ (fixed number of node inputs: 2,3,4). See fig.2.

After the initiation damage may fade out (which is an ordered behaviour) and create the left peak on above distribution, or the damage may reach an equilibrium level which is a chaotic behaviour and creates the right peak (large percolating cluster). The fraction of damage initiation events which exhibit such chaotic behaviour is treated as degree $c$ of chaos. 
This is the first parameter with which maturation is observed. The `chaotic' cases from right peak reach the equilibrium level near or far from a theoretical value expected for a large network in Derrida annealed model. This is the second relation with which maturation is observed. Damage $L$ observed on the network external outputs is an effect of damage process inside the network described by $d$, however a simple relation $L=d*m$ appears only for large networks and this is the third relation with which maturation is observed. Each of these three relations seems to be differently dependent on the observed parameters in detail but generally they give similar conclusions. Differences in these relations' behaviour suggest that these are at least three different mechanisms. 

Two practical criterion types of maturity threshold were discussed: the first one based on the degree of nearness to the known asymptote - e.g. level of damage size equilibrium for a large network, and the second one based on the observed phenomenon of zero appearance between peaks in distribution of frequency of damage size reached after initiation.  
The former one has a clear interpretation, corresponding well to the disappearance of small network effects, however, it does not well correspond to the qualitative view of maturation of base distribution depicted in fig.2.  
The best found candidate seems to be the criterion using appearance of zero, however, it depends a little on event number and works correctly only for networks with feedbacks. For a network without feedbacks (which is a strange case) it can be used with approximation.
However, both found types of criterion -  first, termed `exactness level', allowing to neglect small network effects, and  second - `zero appearance' should be treated as symbols of wide set of more exactly defined criterions based on these phenomena. They can be defined for particular  necessities and are not developed here.

Maturation of network during growth concerns a few (at least three) different phenomena which in different ways depend on the network $type$ and parameters $s$ and $K$. Most of them wait a theoretical description and explanation.  In such a case one criterion for maturation threshold is a simplification which cannot work well for all cases. The appearance of zero is shown in fig.3-5, 7 using lines connecting points for different network types. These lines are generally sloped down to the right which means that this criterion decides  the threshold earlier for less chaotic network types than the first type of criteria. However, such decision corresponds well to the wide qualitative view, which, unfortunately, is not well defined. Also unfortunately, the less chaotic network types $sh$ and $sf$ are the most attractive for applications. This criterion corresponds well to the term `complex network', which is very often used, typically in such an intuitive meaning, but not well defined. It is obvious that such a definition, concerning only chaotic network, is not applicable for all cases, but, I hope, it is useful and may be checked in practice.

In comparison to the main stream of investigation of transition to chaos which concerns the exact `edge of chaos' ($K\leq 2$, $s=2$) and number and length of attractors this approach moves into area known as chaotic ($K\geq 2$, $s\geq 2$ without the case $s,K=2,2$) and reaches much higher network sizes ($N=4000$ even for $s,K=4,4$ and 16,3). In this area other transition to `matured' chaos and its threshold is investigated.
In the main investigated distribution \\ $P(d$ or $L |N,K,s,type)$  the shape containing two peaks and minimum up to zero between them, width, interpretation and position of these peaks are explained, however lots of details still await explanation.

Also in comparison to the main stream, living systems are shifted follow Ref.\citep{agec} to `matured chaos' onto opposite shore of Kauffman's liquid region near `edge of chaos'. Role of negative feedbacks is stressed in stability of living objects.

By the way some simplifications were used and checked which the reader may be interested in. E.g. in additions to the earlier described algorithm a process which I called reversed-annealed approximation was used to check structure influence: the node connection in network was kept (we do not use a particular function) but all node states were randomly changed.  

\section*{Acknowledgments}
Research financed by Polish government as grant 3 T11F 035 30.


\end{document}